\documentclass[british]{ws-mpla}
\usepackage[T1]{fontenc}
\usepackage[latin9]{inputenc}
\usepackage{color}
\usepackage{amsbsy}
\usepackage{graphicx}
\usepackage{esint}
\usepackage{babel}

\usepackage[super,compress]{cite}
\usepackage[breaklinks]{hyperref}  
\hypersetup{colorlinks,urlcolor=black,citecolor=blue,linkcolor=blue,filecolor=black}

\providecommand{\tabularnewline}{\\}

\global\long\def\mpl{m_{\mathrm{Pl}}}

\global\long\def\k{\mpl^{-2}}

\global\long\def\s{\varphi}

\global\long\def\a{A}

\global\long\def\ii{\Vert}

\global\long\def\ev{\lambda}

\global\long\def\ef{\Gamma}

\global\long\def\sv#1{\sigma_{#1}}

\global\long\def\st#1{\sigma_{#1}}

\global\long\def\x{x}

\global\long\def\y{y}

\global\long\def\z{z}

\begin{document}

\markboth{Mindaugas Kar\v{c}iauskas}
{Dynamical Analysis of Anisotropic Inflation}

\catchline{}{}{}{}{}

\title{DYNAMICAL ANALYSIS OF ANISOTROPIC INFLATION}

\author{\footnotesize MINDAUGAS KAR\v{C}IAUSKAS}

\address{Theory Center, IPNS, KEK,\\
Tsukuba 305-0801, Japan\\
and\\ 
University of Jyvaskyla, Department of Physics, \\ 
P.O.Box 35 (YFL), FI-40014 University of Jyv\"{a}skyl\"{a}, Finland\\
mindaugas.m.karciauskas@jyu.fi}

\maketitle

\pub{Received 16 February 2016}{Revised (Day Month Year)}

\begin{abstract}
The inflaton coupling to a vector field via the $f\left(\s\right)^{2}F_{\mu\nu}F^{\mu\nu}$
term is used in several contexts in the literature, such as to generate
primordial magnetic fields, to produce statistically anisotropic curvature
perturbation, to support anisotropic inflation, and to circumvent the
$\eta$-problem. In this work I perform dynamical analysis of this system
allowing for the most general Bianchi I initial conditions. I also confirm
the stability of attractor fixed points along phase-space directions
that had not been investigated before. 
\end{abstract}
\maketitle

\section{Introduction}

A vector field with a varying kinetic function of the form $f^{2}\left(\s\right)F_{\mu\nu}F^{\mu\nu}$,
where $F_{\mu\nu}\equiv\partial_{\mu}\a_{\nu}-\partial_{\nu}\a_{\mu}$,
is often being used in the literature on cosmological inflation for
several different purposes. This type of kinetic function breaks the
conformal invariance of the vector field without introducing instabilities.\cite{Carroll_etal(2009)Instabilities}

By identifying $\a_{\mu}$ with electromagnetic potential such a term
for the first time was studied as a possible source of primordial
magnetic fields\cite{Ratra(1992)} and is still being used for that
purpose (see e.g. Ref.~\refcite{Green(2015)PMFs}). More generally,
a vector field can affect cosmological perturbations or large scale
dynamics of the universe during inflation. In Refs.~\refcite{Dimopoulos2007}--\refcite{Dimopoulos_us(2009)_fF2_PRL}
a massive vector field with $f^{2}F^{2}$ kinetic term plays a role
of the curvaton field. In Refs.~\refcite{Yokoyama_Soda(2008)} and \refcite{Dimopoulos_etal_anisotropy(2008)}
it was demonstrated that the contribution of the vector field to the
primordial curvature perturbation can generate a new observable signature:
 statistical anisotropy. Moreover, it can affect large scale dynamics
by supporting anisotropic inflation.\cite{Kanno_etal(2008)anisInfl}
Finally, if $\s$ is an inflaton, this type of kinetic term can help
to overcome the $\eta$-problem of inflation.\cite{Dimopoulos_Wagstaff(2010)BackReact,Dimopoulos_etal(2012)vFd-eta}

Generally vector field contribution to the curvature perturbation
introduces quadrupolar angular dependence of the power spectrum. In
Ref.~\refcite{Bartolo(2013)fF2} (see also \refcite{Lyth(2013)f2F2}--\refcite{Naruko(2014)anis})
it was shown that observational constraints on such modulation put
very stringent bounds on vector field models. This bound, however,
was applied for a particular model of Ref.~\refcite{Watanabe_Kanno_Soda(2009)}.
As it is shown in later sections, a much larger parameter space might
be available for model building.

In this work I perform a classical dynamical analysis of the system
with a scalar and a vector field which interacts via a $f^{2}\left(\s\right)F_{\mu\nu}F^{\mu\nu}$
coupling. A similar analysis has been performed in Ref.~\refcite{Kanno_etal(2010)attractor}
and much more generally in Ref.~\refcite{Dimopoulos_Wagstaff(2010)BackReact}.
 It has been also applied in many other contexts, for example a system of 
non-Abelian vector fields\cite{Yamamoto(2012)f2F2}, a vector field plus a two-form field\cite{Ito(2015)anisInfl} or DBI type action\cite{Do(2011)f2F2}. Building on the 
former two works, especially
Ref.~\refcite{Dimopoulos_Wagstaff(2010)BackReact}, I extend their treatment
by including more general initial conditions and by addressing several
issues which have been overlooked in previous works. I also provide exact
analytic expressions to determine the asymptotic behaviour of this
dynamical system.

A vector field introduces anisotropic stress. If it contributes to
the expansion of the universe, the expansion will be anisotropic.
As the (homogeneous, space-like) vector field picks out only one preferred
direction in space, we can assume that the anisotropy of the expansion
will be of the axisymmetric Bianchi type I. Moreover, the preferred
axis of such anisotropy will be aligned with the direction of the vector
field. These two constraints were imposed in previous works on a single
vector field. Even if such a configuration is self consistent, initial
conditions can be much more general. In this work I relax both the
constraints and confirm that the axial symmetry as well as the
alignment are achieved dynamically if the vector field does not decay.

In cases where the vector field is absent a number of no-hair theorems
have been proven, which state that the future attractor of any spatially
homogeneous but anisotropic universe (with possible exception of Bianchi
IX\footnote{An analysis of Bianchi IX type model with possible inflationary 
solutions is discussed in Ref.~\refcite{Sundell(2015)anis}.}) is the isotropic FRW 
universe if the energy density is dominated by the potential energy of a scalar field.\cite{Wald(1983)anisBckgr},\footnote{More generally ,this statement is valid for any perfect fluid with
an equation of state $-1\le w<-1/3$, where $P=w\rho$. 
} See Ref.~\refcite{Brandenberger(2016)nohair} for a recent review
and Ref.~\refcite{Kleban(2016)nohair} for the most recent work on no-hair
theorems. The presence of a vector field introduces anisotropic
stress into the energy-momentum tensor. For some parameter values,
as we will see later in this letter, the anisotropic stress does not redshift
away with the expansion of the universe, therefore violating some of
the assumptions underlying no-hair theorems. In particular, due to
the non-vanishing homogeneous vector field, the isotropic FRW universe
is no longer an attractor but rather an axisymmetric Bianchi I universe
is the attractor. This is indeed demonstrated to be the case starting from Bianchi
type II and III as well as Kantowski-Sachs metrics in Ref.~\refcite{Hervik_etal(2011)BIattract}.
However, these authors assumed from the onset an alignment of the
vector field with one of the preferred directions of the metric. More
generally, the authors of Ref.~\refcite{Maleknejad(2012)f2F2} derived
an expression which relates the spatial anisotropy to a general form
of anisotropic stress, assuming an inflating universe. In this letter,
I show by a concrete example how a generic Bianchi I space-time dynamically
aligns itself with a direction of the vector field and two perpendicular
axes of the Bianchi I metric isotropise. For constant model parameters
this attractor is reached from the most general initial conditions.

Even if one does restrict initial conditions to the axial symmetry
and the orientation of that axis along the homogeneous vector field,
effectively excluding some dynamical degrees of freedom, it still
needs to be recognised that to ascertain the stability of a particular
configuration, the stability of those degrees of freedom must be demonstrated.
This is particularly relevant when dealing with inflation models where
any fixed point is constantly being perturbed by quantum stochastic
noise terms. Moreover, those degrees of freedom are more important
in the present setup because anisotropic expansion couples modes of
different spin. This is in contrast to FRW metric, where scalar, vector
and tensor modes evolve independently. To account for this enlarged
dynamical phase space six equations are added to the ones considered
in previous works on anisotropic inflation with $f\left(\s\right)^{2}F_{\mu\nu}F^{\mu\nu}$
term. 

\section{Dynamical System Analysis}

\subsection{The action and the background metric\label{sub:Action-Metric}}

Consider a classical dynamical system described by the following action
\begin{equation}
S=\int\mathrm{d}^{4}x\sqrt{-g}\left[\frac{1}{2}\mpl^{2}R-\frac{1}{2}\partial^{\mu}\s\partial_{\mu}\s-V\left(\s\right)-\frac{1}{4}f^{2}\left(\s\right)F^{\mu\nu}F_{\mu\nu}\right],\label{action}
\end{equation}
where the first term is the Einstein-Hilbert term, $\s$ is the scalar
field with a potential $V\left(\s\right)$, $F_{\mu\nu}$ is the vector
field strength tensor, and the kinetic function $f\left(\s\right)$
is modulated by the scalar field. The presence of the vector field
provides anisotropic stress, which likely makes the spacetime metric
anisotropic. The simplest anisotropic metric consistent
with the symmetries of the energy-momentum tensor in this setup is
the Bianchi type I. In some of the computations below I will use
part of the formalism of Ref.~\refcite{Pereira_etal(2007)anisoBckgr}
and their notation. In that work the Bianchi I metric is written
\begin{equation}
\mathrm{d}s^{2}=-\mathrm{d}t^{2}+\sum_{i=1}^{3}S_{i}^{2}\left(t\right)\left(\mathrm{d}x_{i}\right)^{2},\label{metric-ini}
\end{equation}
where $S_{i}\left(t\right)$ are scale factors along three spatial
directions. To write the above expression, the orientation of the coordinate
system is fixed to make the metric diagonal.

We can factor out the average scale factor from the spatial part of
the metric
\begin{equation}
a\left(t\right)=\left[S_{1}\left(t\right)S_{2}\left(t\right)S_{3}\left(t\right)\right]^{1/3}.\label{a-def}
\end{equation}
In that case the anisotropic part of that metric can be written as
\begin{equation}
\gamma_{ij}\left(t\right)=\mathrm{e}^{2\beta_{i}\left(t\right)}\delta_{ij}\label{gamma-beta}
\end{equation}
where $\sum_{i}^{3}\beta_{i}\left(t\right)=0$ and no summation implied.
By rescaling the time coordinate as $\mathrm{d}\tau=\mathrm{d}t/a$
where $\tau$ is the so called conformal time, the metric in Eq.~(\ref{metric-ini})
becomes 
\begin{equation}
\mathrm{d}s^{2}=a^{2}\left(\tau\right)\left[-\mathrm{d}\tau^{2}+\gamma_{ij}\left(\tau\right)\mathrm{d}x^{i}\mathrm{d}x^{j}\right].
\end{equation}
The volume expansion rate and shear tensor can then be given as
\begin{equation}
\mathcal{H}\equiv\frac{a'}{a}\quad\mathrm{and}\quad\sigma_{ij}\equiv\frac{1}{2}\gamma_{ij}',
\end{equation}
where primes denotes differentiation with respect to $\tau$.

It is also convenient to define a volume expansion rate with respect
to the cosmic time $t$, which is given by $H\equiv\dot{a}/a$, where
$a=a\left(t\right)$ is defined in Eq.~(\ref{a-def}) and the dot represents
a time derivative with respect to $t$. One can easily check that
$\mathcal{H}=aH$.

\subsection{Einstein equations}

To write the Einstein equations, it is customary to decompose the energy-momentum
tensor with respect to the average velocity vector $u^{\mu}$, which
is a time-like four vector field normalised $u_{\mu}u^{\mu}=-1$.
Since we are interested in background quantities of non-tilted fluids\footnote{The flow lines of tilted fluids are not orthogonal to the surfaces of homogeneity. See, e.g., Refs.~\refcite{Ellis(1971)} and \refcite{Ellis-DynSys}.},
the energy flux relative to $u^{\mu}$ can be neglected and the general
energy-momentum tensor can be written as
\begin{equation}
T_{\mu\nu}=\rho u_{\mu}u_{\nu}+P\left(g_{\mu\nu}+u_{\mu}u_{\nu}\right)+\pi_{\mu\nu},\label{Tmn-gen}
\end{equation}
where $\rho$ is the total energy density, $P$ is the total pressure,
and $\pi_{\mu\nu}$ is anisotropic stress satisfying $\pi_{\mu\nu}u^{\mu}=0$
and $\pi_{\mu}^{\mu}=0$. 

The scalar and vector field energy-momentum tensors can be obtained
by varying the action in Eq.~(\ref{action}) with respect to the
metric $g_{\mu\nu}$
\begin{eqnarray}
T_{\mu}^{\left(\s\right)\nu} & = & \partial_{\mu}\s\partial^{\nu}\s-\delta_{\mu}^{\nu}\left(\frac{1}{2}\partial_{\rho}\s\partial^{\rho}\s+V\right),\\
T_{\mu}^{\left(\a\right)\nu} & = & -f^{2}\left(\frac{1}{4}\delta_{\mu}^{\nu}F_{\rho\sigma}F^{\rho\sigma}-F_{\mu\rho}F^{\nu\rho}\right).
\end{eqnarray}
Comparing these two expressions with Eq.~(\ref{Tmn-gen}), we easily
find the energy density, pressure, and anisotropic stress for each
component
\begin{eqnarray}
\rho_{\s} & = & \frac{1}{2}a^{-2}\left(\s'\right)^{2}+V,\quad P_{\s}=\frac{1}{2}a^{-2}\left(\s'\right)^{2}-V\label{rho-phi}\\
\rho_{\a} & = & \frac{1}{2}f^{2}\gamma^{ij}\frac{\a_{i}'\a_{j}'}{a^{4}},\quad P_{\a}=\frac{1}{3}\rho_{\a},\quad\pi_{j}^{i}=-f^{2}\left(\delta_{j}^{l}\gamma^{ik}-\frac{1}{3}\delta_{j}^{i}\gamma^{kl}\right)\frac{\a_{l}'\a_{k}'}{a^{4}}\label{pi-A}
\end{eqnarray}

The general expressions of the Einstein's field equations with anisotropic
energy-momentum tensor can be found in the Appendix A of Ref.~\refcite{Pereira_etal(2007)anisoBckgr}.
Plugging Eqs.~(\ref{rho-phi}) and (\ref{pi-A}) into those expressions,
we get
\begin{eqnarray}
\mathcal{H}^{2} & = & \frac{1}{3\mpl^{2}}\left[\frac{1}{2}\s'^{2}+a^{2}V+\frac{1}{2}f^{2}\frac{\gamma^{ij}\a_{i}'\a_{j}'}{a^{2}}\right]+\frac{1}{6}\sigma^{2},\label{H2}\\
\mathcal{H}' & = & -\frac{1}{3\mpl^{2}}\left[\s'^{2}-a^{2}V+\frac{1}{2}f^{2}\frac{\gamma^{ij}\a_{i}'\a_{j}'}{a^{2}}\right]-\frac{1}{3}\sigma^{2},\label{dH}\\
\left(\sigma_{j}^{i}\right)' & = & -2\mathcal{H}\sigma_{j}^{i}+\k\left(\frac{1}{3}\delta_{j}^{i}\gamma^{kl}-\delta_{j}^{l}\gamma^{ik}\right)f^{2}\frac{\a_{l}'\a_{k}'}{a^{2}}.\label{s-eqn}
\end{eqnarray}

\subsection{Decomposition of the shear tensor}

$\sigma_{ij}$ is a symmetric, trace free tensor. In the coordinate
system in Eq.~(\ref{metric-ini}) $\sigma_{ij}$ has only two components:
the off-diagonal components of $\sigma_{ij}$ vanish and the remaining
two are given by the time derivative of functions $\beta_{i}\left(\tau\right)$
in Eq.~(\ref{gamma-beta}). In an arbitrary coordinate system, however,
three additional parameters are needed to specify three angles of
$\sigma_{ij}$ with respect to that coordinate system. Thus in total
there are five independent components.

In our case, the only source of the shear tensor $\sigma_{ij}$ is
anisotropic stress generated by the presence of the vector field.
Thus, to study the evolution of $\sigma_{ij}$ it is convenient to
choose a reference frame related to the vector field $\a_{i}$.
For that purpose we introduce a unit vector $\hat{z}_{i}$ such that
\begin{equation}
\hat{z}_{i}\equiv\frac{\a_{i}'}{\a'}\quad\mathrm{and}\quad\a'\equiv\sqrt{\gamma^{ij}\a_{i}'\a_{j}'}\label{zi-dA-def}
\end{equation}
where $\a'$ is the magnitude of the $\a_{i}'$ vector. The shear tensor
$\sigma_{ij}$ can be decomposed into components relative to the orthogonal
basis $\left(\hat{\boldsymbol{z}},\boldsymbol{e}^{1},\boldsymbol{e}^{2}\right)$,
where
\begin{equation}
\gamma^{ij}\hat{z}_{i}e_{j}^{a}=0\quad\mathrm{for}\quad a=1,2.\label{ei-def}
\end{equation}
Adapting the expression in Eq.~(3.1) of Ref.~\refcite{Pereira_etal(2007)anisoBckgr},
we can write
\begin{equation}
\sigma_{ij}=\frac{3}{2}\left(\hat{z}_{i}\hat{z}_{j}-\frac{1}{3}\gamma_{ij}\right)\sigma_{\ii}+2\sv a\hat{z}_{\left(i\right.}e_{\left.j\right)}^{a}+\st{\lambda}\varepsilon_{ij}^{\lambda}.\label{s-decomp}
\end{equation}
In this expression summation over the repeated indices $a=1,2$ and
$\lambda=+,\times$ is implied and parentheses in the indices denote
symmetrization $a_{\left(i\right.}b_{\left.j\right)}\equiv\left(a_{i}b_{j}+a_{j}b_{i}\right)/2$.
The polarization tensor $\varepsilon_{ij}^{\lambda}$ is defined as
\begin{equation}
\varepsilon_{ij}^{\lambda}=\frac{1}{\sqrt{2}}\left(e_{i}^{1}e_{j}^{1}-e_{i}^{2}e_{j}^{2}\right)\delta_{+}^{\lambda}+\frac{1}{\sqrt{2}}\left(e_{i}^{1}e_{j}^{2}+e_{i}^{2}e_{j}^{1}\right)\delta_{\times}^{\lambda}.
\end{equation}
I will use the notation where $\sv a$ with a Latin letter index from
the beginning of the alphabet denotes the two vector components of $\sigma_{ij}$,
while Greek letter index, as in $\st{\lambda}$, denotes the two tensor
components.

As noted in the Introduction, none of the preferred
directions of $\sigma_{ij}$ are constrained to be aligned with the direction of $\hat{z}_{i}$
initially. Neither $\sigma_{ij}$ is assumed to be axisymmetric. Such
an alignment and symmetry correspond to $\sigma_{\ii}\ne0$ and $\sv a=\st{\lambda}=0$
for all $a$ and $\lambda$. Instead, we find this configurations
to be achieved dynamically in cases where the vector field does not
vanish. 

To derive equations of motion for $\sigma_{\ii}$, $\sv a$, and $\st{\lambda}$,
we need to compute time derivatives of basis vectors $\boldsymbol{e}^{a}$
first. This can be done after fixing the remaining freedom in the
definition of $\boldsymbol{e}^{a}$. So far these vectors were defined
up to a rotation around $\hat{\boldsymbol{z}}$. Their definition
is completed by imposing $\mathcal{R}_{\left[ab\right]}=0,$ where
$\mathcal{R}$ is defined as $\left(e_{a}^{i}\right)'=\sum_{b}\mathcal{R}_{ab}e_{b}^{i}$
and $\left[a,b\right]\equiv\left(ab-ba\right)/2$. Since the $\boldsymbol{e}^{a}$
are orthonormal by definition, they satisfy $\left(e_{a}^{i}e_{i}^{b}\right)'=0$,
which allows us to write $\left(e_{a}^{i}\right)'=\left(2\sigma_{ij}-\sigma_{lj}e_{b}^{l}e_{i}^{b}\right)e_{a}^{j}$.
Using this equation and plugging Eq.~(\ref{s-decomp}) into (\ref{s-eqn}),
we obtain equations of motion for each component of $\sigma_{ij}$
\begin{eqnarray}
\sigma_{\ii}'+2\mathcal{H}\sigma_{\ii} & = & -2\sigma_{V}^{2}-\frac{2}{3}\k f^{2}\frac{\a'^{2}}{a^{2}}\label{EoM-sii}\\
\sv a'+2\mathcal{H}\sv a & = & \frac{1}{2}\sv a\sigma_{\ii}-\sv{}^{b}\st{\lambda}\mathcal{M}_{ab}^{\lambda}\\
\st{\lambda}'+2\mathcal{H}\st{\lambda} & = & 2\sv{}^{a}\sv{}^{b}\mathcal{M}_{ab}^{\lambda}\label{EoM-sT}
\end{eqnarray}
where $\sigma_{V}^{2}\equiv\sv{}^{a}\sv a$ and $\mathcal{M}_{ab}^{\lambda}\equiv\varepsilon_{ij}^{\lambda}e_{a}^{i}e_{b}^{j}$.
To derive these equations, $\hat{z}^{i}\a_{i}'=\a'$ and $e_{a}^{i}\a_{i}'=0$
were also used.

\subsection{Scalar and vector field equations of motion}

To close the system of equations we must specify equations of motion
for the fields $\s$ and $\a_{i}$. The former can be obtained by varying
the action in Eq.~(\ref{action}) with respect to $\s$, which gives
\begin{equation}
\s''+2\mathcal{H}\s'+a^{2}V_{,\s}=\frac{1}{2}a^{-2}f_{,\s}^{2}\a'^{2},\label{EoM-sFd}
\end{equation}
where $V_{,\s}$ denotes the derivative with respect to a scalar field
$V_{,\s}\equiv\mathrm{d}V/\mathrm{d}\s$ and similarly for $f_{,\s}^{2}\equiv\mathrm{d}\left(f^{2}\right)/\mathrm{d}\s$.
The equation for the vector field can be obtained by varying the same
action with respect to $\a_{i}'$. In this case we find
\begin{equation}
\a_{i}''+2\left(\frac{f'}{f}-\sigma_{\ii}\right)\a_{i}'=2\sv ae_{i}^{a}\a'.\label{EoM-vFd}
\end{equation}


\subsection{Expansion-normalised variables}

Equations (\ref{EoM-sii})-(\ref{EoM-vFd})
together with the energy constraint in Eq.~(\ref{H2}) form a closed
system of coupled dynamical equations. Equation~(\ref{dH}) will be useful
to determine which solutions correspond to an accelerated expansion,
i.e. inflationary solutions. However, instead of analysing these equations
in the form given above, it is more convenient to rewrite them in terms
of expansion-normalised variables.\cite{Ellis-DynSys} For this purpose
we define 
\begin{eqnarray}
\Sigma_{\ii} & \equiv & \frac{\sigma_{\ii}}{\mathcal{H}},\quad\Sigma_{a}\equiv\frac{\sigma_{a}}{\mathcal{H}},\quad\Sigma_{\lambda}\equiv\frac{\sigma_{\lambda}}{\mathcal{H}},\\
\x & \equiv & \frac{\s'}{\sqrt{6}\mpl\mathcal{H}},\quad\y\equiv\frac{a\sqrt{V}}{\sqrt{3}\mpl\mathcal{H}},\quad\z\equiv\frac{f\a'/a}{\sqrt{6}\mpl\mathcal{H}}.\label{x-def}
\end{eqnarray}
Hence each of the solutions $\boldsymbol{X}\equiv\left(\Sigma_{\ii},\Sigma_{a},\Sigma_{\lambda},\x,\y,\z\right)$
corresponds to a family of solutions in physical space, given by $\left(H,\boldsymbol{X}\right)$.
In addition to making expressions more compact, this normalisation
decouples the Raychaudhury Eq.~(\ref{dH}) from the rest of the equations
and leads to a bounded dynamical system. consequently, the closed system of
dynamical equations become
\begin{eqnarray}
\epsilon_{H} & = & 3\left(\x^{2}+\Sigma^{2}\right)+2\z^{2}\label{eH-eqn}\\
1 & = & \Sigma^{2}+\x^{2}+\y^{2}+\z^{2}\label{constr}\\
\frac{\mathrm{d}\Sigma_{\ii}}{\mathrm{d}N} & = & \Sigma_{\ii}\left(\epsilon_{H}-3\right)-2\Sigma_{V}^{2}-4\z^{2}\label{EoM-Sii}\\
\frac{\mathrm{d}\Sigma_{b}}{\mathrm{d}N} & = & \Sigma_{b}\left(\epsilon_{H}-3\right)+\frac{1}{2}\Sigma_{b}\Sigma_{\ii}-\Sigma^{a}\Sigma_{\lambda}\mathcal{M}_{ab}^{\lambda}\\
\frac{\mathrm{d}\Sigma_{\lambda}}{\mathrm{d}N} & = & \Sigma_{\lambda}\left(\epsilon_{H}-3\right)+2\Sigma^{a}\Sigma^{b}\mathcal{M}_{ab}^{\lambda}\\
\frac{\mathrm{d}\x}{\mathrm{d}N} & = & \x\left(\epsilon_{H}-3\right)-\ev\y^{2}+\ef\z^{2}\\
\frac{\mathrm{d}\z}{\mathrm{d}N} & = & \z\left(\epsilon_{H}-3\right)+\z\left(\Sigma_{\ii}+1-\ef\x\right)\label{EoM-z}
\end{eqnarray}
where the slow-roll parameter is $\epsilon_{H}\equiv-\dot{H}/H^{2}$.
In these equations the time variable is replaced by the number of
e-folds of expansion $N\equiv\ln\left(\frac{a}{a_{*}}\right),$ where
$a_{*}$ denotes an initial scale factor. $\Sigma_{V}^{2}$ and $\Sigma^{2}$
in the above equations are defined as
\begin{equation}
\Sigma_{V}^{2}\equiv\Sigma_{a}\Sigma^{a}\quad\mathrm{and}\quad\Sigma^{2}\equiv\frac{1}{6}\left(\frac{3}{2}\Sigma_{\ii}^{2}+2\Sigma_{a}\Sigma^{a}+\Sigma_{\lambda}\Sigma^{\lambda}\right).\label{Sig2-def}
\end{equation}

$\ev$ and $\ef$ are the free parameters of the model. Their values
determine the asymptotic behaviour of the system. These parameters
are given by
\begin{equation}
\ev\equiv\sqrt{\frac{3}{2}}\mpl\frac{V_{,\s}}{V}\quad\mathrm{and}\quad\ef\equiv\sqrt{6}\mpl\frac{f_{,\s}}{f}.\label{eV-ef-def}
\end{equation}
Note that $\ev$ is related to the standard slow-roll parameter $\epsilon_{V}$
by
\begin{equation}
\epsilon_{V}\equiv\frac{\mpl^{2}}{2}\left(\frac{V_{,\s}}{V}\right)^{2}=\frac{\ev^{2}}{3}.\label{epsilonV-def}
\end{equation}

\subsection{Fixed points}

It is impossible to solve non-linear coupled Eqs.~(\ref{EoM-Sii})--(\ref{EoM-z})
in full generality exactly. However, it is enough to determine only
the asymptotic behaviour of the solutions. This can be done by performing
the dynamical analysis, that is, by finding fixed points and
checking their stability. If a given fixed point is found to
be a unique attractor in the sense that no other attractor or periodic
orbit exist simultaneously, the system will asymptotically evolve
to that fixed point starting from any initial conditions.

Fixed points (or fixed points) correspond to the state of the
system which does not evolve in time. Hence, we want to find those
points in the compact phase space spanned by variables $\boldsymbol{X}$,
where l.h.s. of Eqs.~(\ref{EoM-Sii})--(\ref{EoM-z}) vanish. Setting
$\mathrm{d}\boldsymbol{X}/\mathrm{d}N=0$, those equations reduce
to a system of algebraic equations. It is much easier to find their
solutions than to solve differential equations.

For the solutions to be physical, they have to satisfy two conditions.
First, they have to be real and they have to satisfy the Hamiltonian
energy constraint in Eq.~(\ref{constr}), that is $\Sigma^{2}+\x^{2}+\z^{2}\le1$.

To find fixed points let us assume the $\ev$ and $\ef$ parameters
to be constant. In that case five distinct fixed points exist
which satisfy the above two conditions. They are listed in Table~1.
For brevity $\Delta$'s in this table are defined as{\small{}
\begin{eqnarray}
\Delta_{\mathrm{I}} & \equiv & 12+\left(\ef+\ev\right)\left(3\ef+\ev\right),\\
\quad\Delta_{\mathrm{II}} & \equiv & 12+\left(\ef+\ev\right)\left(3\ef-\ev\right),\\
\quad\Delta_{\mathrm{III}} & \equiv & -6+\ev\left(\ef+\ev\right).
\end{eqnarray}
}{\small \par}

\begin{table}
\begin{centering}
{Table 1. ~ Five fixed points of the system of equations (\ref{EoM-Sii})--(\ref{EoM-z}).}\\
~
\par\end{centering}

\begin{centering}
\begin{tabular}{cccccc}
\hline 
Fixed Point & $\Sigma_{\ii}$ & $\Sigma_{a}$ & $\Sigma_{\lambda}$ & $\x$ & $\z$\tabularnewline
\hline 
~ ~$\mathcal{K}_{\pm}$~ ~ & $\Sigma_{\ii}$ & ~ 0 ~  & ~ 0 ~ & $\pm\sqrt{1-\Sigma^{2}}$ & 0\tabularnewline
$\mathcal{S}$ & 0 & 0 & 0 & $-\frac{\ev}{3}$ & 0\tabularnewline
$\mathcal{V}_{\pm}$ & ~ $-\frac{4\Delta_{\mathrm{III}}}{\Delta_{\mathrm{I}}}$ ~ & 0 & 0 & ~ $-\frac{6\left(\ef+\ev\right)}{\Delta_{\mathrm{I}}}$ ~ & ~ $\pm\frac{\sqrt{3\Delta_{\mathrm{II}}\Delta_{\mathrm{III}}}}{\Delta_{\mathrm{I}}}$
~\tabularnewline
\hline 
\end{tabular}
\par\end{centering}

\protect\caption{}
\end{table}

The first two fixed ``points'' $\mathcal{K}_{\pm}$ are not
isolated points but rather a set of points which form a hypersphere
given by $\Sigma_{\mathcal{K}}^{2}+\x_{\mathcal{K}}^{2}=1$ and $\z_{\mathcal{K}}=0$.
Subscripts will be used to denote values of dynamical variables in
a given fixed point. The potential energy of the scalar field
vanishes on $\mathcal{K}_{\pm}$, i.e. $\y_{\mathcal{K}}=0$. In
Ref.~\refcite{Dimopoulos_Wagstaff(2010)BackReact} this
hypersphere of fixed points was called the anisotropic kination solution ($\mathcal{AKS}$).

The third fixed point $\mathcal{S}$ corresponds to the case
in which both the anisotropy and the vector field energy density
vanishes, so that $\Sigma_{\mathcal{S}}=0$ and $\z_{\mathcal{S}}=0$, and
the universe is dominated by the scalar field. Its (expansion rate
normalised) kinetic energy is given by $\x_{\mathcal{S}}=-\ev/3$
and potential energy by $\y_{\mathcal{S}}=\sqrt{1-\x_{\mathcal{S}}^{2}}$.

The last two fixed points, denoted by $\mathcal{V}_{\pm}$,
are the most interesting ones where the vector field and spatial anisotropy
do not vanish, i.e., $\z_{\mathcal{V}}=\pm\sqrt{3\Delta_{\mathrm{II}}\Delta_{\mathrm{III}}}/\Delta_{\mathrm{I}}$
and $\Sigma_{\mathcal{V}}=-2\Delta_{\mathrm{III}}/\Delta_{\mathrm{I}}$.


\subsection{Stability analysis}

The asymptotic behaviour of the system depends on the stability of
fixed points in Table~1. At linear level the stability can
be determined by perturbing all the dynamical degrees of freedom as
$\boldsymbol{X}\rightarrow\boldsymbol{X}+\delta\boldsymbol{X}$, where
$\boldsymbol{X}$ on the right-hand side corresponds to the unperturbed values
and $\delta\boldsymbol{X}$ are small perturbations. Plugging this
expression into Eqs.~(\ref{eH-eqn})--(\ref{EoM-z}) we keep only terms up
to the first order in $\delta\boldsymbol{X}$. 

It is important to note that $\delta\boldsymbol{X}$ also includes
the vector perturbation in the directions perpendicular to the homogeneous
value $\hat{z}_{i}$, i.e.,
\begin{equation}
\delta\z=\delta\z_{\ii}\hat{z}_{i}+\delta\z_{a}e_{i}^{a}.
\end{equation}
The stability of fixed points in the $e_{i}^{a}$ directions
has not been checked. However, the complete analysis requires dynamical
equations to be perturbed along these directions too, even if initial
conditions are constrained to be axisymmetric Bianchi I along $\hat{z}_{i}$.
The same comment applies to perturbations along $\Sigma_{a}$ and
$\Sigma_{\lambda}$ directions. Most of the analyses examined only
the behaviour of the $\delta\Sigma_{\ii}$ perturbation.

As homogeneous equations for the vector field along $e_{i}^{a}$ do
not exist, linearised equations of motion for $\delta\z_{\ii}$ and
$\delta\z_{a}$ must be derived from Eq.~(\ref{EoM-vFd}) directly.
The full system of linearised dynamical equations are provided in
the Appendix. They can be written in symbolic form as
\begin{equation}
\frac{\mathrm{d}}{\mathrm{d}N}\left(\delta\Sigma_{\ii}\ldots\delta z_{2}\right)^{\mathrm{T}}=\mathcal{M}\left(\delta\Sigma_{\ii}\ldots\delta z_{2}\right)^{\mathrm{T}},\label{diff-eqns-formal}
\end{equation}
where $\mathcal{M}\left(\ev,\ef\right)$ is a $9\times9$ matrix composed
of fixed point values given in Table~1. As this is a system
of linear first order differential equations, the formal solution can
be written as $\left(\delta\Sigma_{\ii}\ldots\delta z_{2}\right)^{\mathrm{T}}=\mathcal{C}\left(\mathrm{e}^{m_{1}N}\ldots\mathrm{e}^{m_{9}N}\right)^{\mathrm{T}}$,
where $\mathcal{C}$ is the $9\times9$ matrix of integration constants
and $m_{i}$'s are the eigenvalues of the matrix $\mathcal{M}$. If
the real part of $m_{i}$ is positive, a small perturbation around
the fixed point in the $v_{i}$ direction grows with time, where
$v_{i}$ is an eigenvector corresponding to the eigenvalue $m_{i}$.
And vice versa, if the real part of $m_{i}$ is negative, the
perturbation decays exponentially. An fixed point is said to
be an attractor if perturbations along all the $v_{i}$ directions
decay. If some of the real parts of the $m_{i}$'s are positive, such
an fixed point is not stable, as any small deviation from that
point along corresponding $v_{i}$ directions will grow. If real parts
of some eigenvalues $m_{i}$ are zero, the stability of the fixed
point cannot be determined by linear analysis. 

We also encounter non-isolated fixed points, i.e., $\mathcal{K}_{\pm}$.
Such points form an $r$-dimensional set, for example, a curve of
fixed points in a $1$-dimensional case. Each of the non-isolated
fixed points necessarily has $r$ vanishing eigenvalues. If
real parts of all other eigenvalues are negative, such a set is an
attractor because all the orbits approach this $r$-dimensional set
as time grows. A much broader and more rigorous discussion of dynamical
systems and their applications in cosmology can be found in Ref.~\refcite{Ellis-DynSys}.

\begin{figure}
\begin{centering}
\includegraphics[scale=0.4]{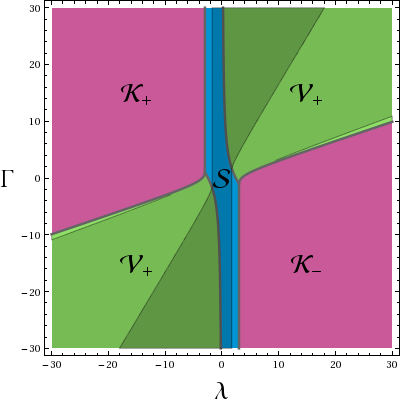}~ \includegraphics[scale=0.4]{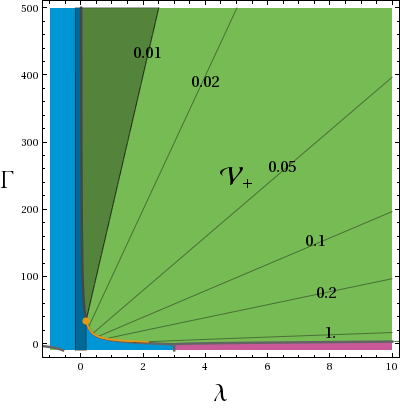}
\par\end{centering}

\protect\caption{
\label{fig:stab-regs}
Stability regions for fixed points of
table~1. Each colour represents a $\left(\protect\ev,\protect\ef\right)$
parameter region where an indicated fixed point is an attractor.
Equations for bifurcation curves separating these regions are given
in Eqs.~(\ref{KV-bndr}), (\ref{SV-bndr}) and (\ref{SK-bndr}).
In the l.h.s. plot darker blue and green regions represent parameter
space where the universe is inflating, i.e. $\epsilon_{H}<1$. The
r.h.s plot is a rescaled version of the l.h.s. one (a symmetric plot
could be drawn for $\protect\ev<0$ and $\protect\ef<0$ values).
In this plot darker region represents the observationally allowed
$\epsilon_{H}<10^{-2}$ range (the origin of this bound is clarified
before Eq.~(\ref{eH-bound})). Contours show constant $\epsilon_{H}$
values. The orange curve in the lower left corner of this plot shows
the trajectory of the model introduced in Ref.~16.
}

\end{figure}

The full analytic expressions of all eigenvalues for each fixed
point are given in the Appendix. Their most important aspects are
also summarised on the left-hand side of Figure~\ref{fig:stab-regs}. Different
colour regions (purple, blue and green) of this plot represent the $\ev$
and $\ef$ parameter values for which a corresponding fixed
point (or a set of points in the case of $\mathcal{K}_{\pm}$) is
an attractor. For example, if the $\ev$ and $\ef$ values fall within
the blue region, the $\mathcal{S}$ fixed point is an attractor.

It is important to notice is that none of the regions overlap.
This means that no two fixed points in Table~1 are attractors
simultaneously. The letter in each coloured region tells which fixed
point is an attractor. For the values of $\ev$ and $\ef$ within
that region, other fixed points are either unstable (some or
all of the real parts of eigenvalues are positive) or they do not exist
(the fixed point is either imaginary or does not satisfy the Hamiltonian
energy constraint). Hence, for fixed values of $\ev$ and $\ef$ the
asymptotic fate of the solution is uniquely determined, irrespective
of initial conditions: it relaxes towards one of fixed points,
depending on which region $\ev$ and $\ef$ fall into.

The only parameter space where the above discussion is not entirely
accurate is represented by narrow green bands between $\mathcal{K}_{\pm}$
and $\mathcal{V}_{+}$ regions. If the $\ev$ and $\ef$ values fall within
those bands, then both $\mathcal{V}_{+}$ as well as $\mathcal{V}_{-}$
fixed points are attractors simultaneously. Otherwise, $\mathcal{V}_{-}$
is a saddle point (real parts of some of $m_{i}$'s are negative and
others are positive ) in the whole $\mathcal{V}_{+}$ region.

Different colour regions are separated by bifurcation curves. They
signify boundaries where the stability of attractor fixed points
changes. These bifurcation curves can be parametrised as follows.
Curves separating $\mathcal{K}_{\pm}$ and $\mathcal{V}_{+}$ regions
(purple and green) are given by a solution of the $\Delta_{\mathrm{II}}=0$
equation
\begin{equation}
\ef=-\frac{\ev}{3}+2\mathrm{sg}\left(\ev\right)\sqrt{\left(\frac{\ev}{3}\right)^{2}-1}\;\mathrm{for}\;\left|\ev\right|\ge3,\label{KV-bndr}
\end{equation}
where $\mathrm{sg}\left(\ev\right)$ is the sign of the parameter
$\ev$. The bifurcation curves separating $\mathcal{S}$ and $\mathcal{V}_{+}$
regions (blue and green) can be parametrised by a solution of $\Delta_{\mathrm{III}}=0$
equation
\begin{equation}
\ef=\frac{6-\ev^{2}}{\ev}\;\mathrm{for}\;\left|\ev\right|\le3.\label{SV-bndr}
\end{equation}
Finally, $\mathcal{S}$ and $\mathcal{K}$ regions (purple and blue)
are separated by lines
\begin{equation}
\left|\ev\right|=3\;\mathrm{for}\;\left|\ef\right|\ge1.\label{SK-bndr}
\end{equation}
For example, if $\left|\ef\right|$ is larger than the values in Eqs.~(\ref{KV-bndr})
and (\ref{SV-bndr}), the only attractor fixed point is $\mathcal{V}_{+}$
(disregarding the $\mathcal{V}_{-}$ bands). Other fixed points
are either unstable or do not exist for these parameter values.

\subsection{Inflationary solutions}

The main goal of this work is to identify the parameter region in the
$\left(\ev,\ef\right)$ plane that accepts inflationary solutions
as their attractor fixed points, i.e. $\epsilon_{H\mathcal{X}}<1$,
where $\mathcal{X}=\mathcal{K}_{\pm},\:\mathcal{S},\:\mathcal{V}_{+}$
denotes the values of $\epsilon_{H}$ in a given attractor. The general
$\epsilon_{H}$ can be written in terms of dimensionless variables
$\Sigma$, $\x$, etc. as in Eq.~(\ref{eH-eqn}). By plugging the
attractor values from Table~1 into this equation, we can find the
parameter space which satisfies the condition $\epsilon_{H\mathcal{X}}<1$.
It is represented by a darker region on the left-hand side of Figure~\ref{fig:stab-regs}.

From this figure we clearly see that the anisotropy and kinetic term
dominated regions (denoted by $\mathcal{K}_{\pm}$) are non-inflationary.
Indeed, the slow-roll parameter $\epsilon_{H\mathcal{K}}=3$ in the
whole parameter space where fixed points $\mathcal{K}_{\pm}$
are attractors. On the other hand, the parameter space with inflationary
attractors overlaps $\mathcal{S}$ and $\mathcal{V}_{+}$ regions.
In the $\mathcal{S}$ part the slow-roll parameter is given by
\begin{equation}
\epsilon_{H\mathcal{S}}=-\ev\x_{\mathcal{S}}=\frac{\ev^{2}}{3}.
\end{equation}

This fixed point corresponds to the case where the vector field
as well as spatial anisotropy vanish. Furthermore, from the expression
of $\x_{\mathcal{S}}$ in Table~1 and the energy constraint in Eq.~(\ref{constr})
we find that the bound $\epsilon_{H\mathcal{S}}<1$ implies $\y_{\mathcal{S}}>\x_{\mathcal{S}}$,
i.e., the universe is dominated by the potential energy of the scalar
field. In this case, the inflating solution is indistinguishable (at
least at the unperturbed level) from a single scalar field inflation
scenario. If $\epsilon_{H\mathcal{S}}$ satisfies a stronger condition $\epsilon_{H\mathcal{S}}\ll1$, then $\y_{\mathcal{S}}\simeq1$ and we recover the standard slow-roll solution ($\mathcal{SSR}$).\cite{Dimopoulos_Wagstaff(2010)BackReact}
Indeed, using Eq.~(\ref{epsilonV-def}) we can easily find that the condition
$\epsilon_{H\mathcal{S}}\ll1$ is equivalent to the standard condition
for the single field slow-roll inflation $\epsilon_{V}\ll1$, where
$\epsilon_{V}$ is defined in Eq.~(\ref{epsilonV-def}).

The darker region also extends into the $\mathcal{V}_{+}$ parameter
space where the vector field and spatial anisotropy do not vanish.
Such fixed points correspond to an inflating universe which circumvents the assumptions
of no-hair theorems. This was first noticed in Ref.~\refcite{Watanabe_Kanno_Soda(2009)}.
The exact expression for the slow-roll parameter in the $\mathcal{V}_{+}$
region can be written as
\begin{equation}
\epsilon_{H\mathcal{V}}=-\ev\x_{\mathcal{V}}=3\left(1-\frac{\Delta_{\mathrm{II}}}{\Delta_{\mathrm{I}}}\right).\label{eHV}
\end{equation}

Even if inflationary attractor solutions can be obtained in the whole
parameter space emphasised by the darker regions on the left-hand side of
Figure~\ref{fig:stab-regs}, the value of $\epsilon_{H}$ when cosmological
scales exit the horizon is bounded to be much smaller. $\epsilon_{H}$
can affect the spectral tilt of the primordial curvature perturbation,
which is tightly constrained by the Planck satellite.\cite{Ade(2015)Planck-infl}
Restricting only to a first order in Hubble flow parameters this bound
is $\epsilon_{H}<0.0068$ at $95\%$ CL. One cannot, however, apply this bound
directly as it was derived for a single scalar field models
of inflation. The presence of a vector field could potentially modify
this bound. However, I assume the bound is not changed drastically and use
\begin{equation}
\epsilon_{H}<10^{-2}.\label{eH-bound}
\end{equation}
The parameter space where this condition is satisfied is shown by
the darker region on the right-hand side of Figure~\ref{fig:stab-regs}. This
plot is a rescaled version of the left-hand side making the relevant
parameter space more pronounced.

As one can read from the figure, observational constraints require
\begin{equation}
\left|\ef\right|\gg\left|\ev\right|\quad\mathrm{and}\quad\ef^{2}\gg1\label{V-infl}
\end{equation}
if inflation is to be realised in the $\mathcal{V}_{+}$ attractor.
In this limit expressions of $\Sigma_{\ii\mathcal{V}}$, $\x_{\mathcal{V}}$
and $\z_{\mathcal{V}}$ can be written in a simplified form as
\begin{equation}
\Sigma_{\ii\mathcal{V}}\simeq-\frac{4}{3}\frac{\ef\ev-6}{\ef^{2}},\quad\x_{\mathcal{V}}\simeq-\frac{2}{\ef},\quad\z_{\mathcal{V}}\simeq\sqrt{\frac{\ef\ev-6}{\ef^{2}}},\label{zV-lim}
\end{equation}
where terms up to first order in $\ev/\ef$ and $1/\ef^{2}$ are displayed.
Note that for $\ef\ev<6$ $\mathcal{V}_{+}$ is not an attractor.
The expression of $\epsilon_{H\mathcal{V}}$ is also simplified in
this region
\begin{equation}
\epsilon_{H\mathcal{V}}\simeq2\frac{\ev}{\ef}.\label{eHV-lim}
\end{equation}
Using the definition of the standard slow-roll parameter $\epsilon_{V}$
in Eq.~(\ref{epsilonV-def}) and the above expression we can write
$\epsilon_{H\mathcal{V}}\simeq2\sqrt{3\epsilon_{V}}/\ef$. It becomes
clear from this result that in the $\mathcal{V}_{+}$ attractor inflationary
solutions are possible even if the scalar field potential is steep,
that is, there is a large parameter space where $\epsilon_{H}\ll1$
for $\epsilon_{V}>1$. 

The possibility of having an inflationary solution even in the steep
region of the scalar field potential was first pointed out in Ref.~\refcite{Dimopoulos_Wagstaff(2010)BackReact}.
Based on this result one may hope that the scalar-gauge field interaction
of the form $f^{2}\left(\s\right)F^{2}$ could be a viable candidate
to solve the infamous $\eta$-problem, which plagues inflation model
building in the framework of supergravity theories of particle physics
(see e.g. Ref.~\refcite{Lyth_Liddle(2009)book} and references therein).
This result suggests that if one includes the gauge sector into inflation
model building, it might be possible to escape the $\eta$-problem
while keeping a generic form of the K\"{a}hler potential. The gauge
kinetic function is an integral part of supergravity theories, where
$f$ is a holomorphic function of scalar fields. By including this sector
it might be possible to achieve dynamical cancellation of large mass
terms. A model along these lines, for example, was constructed in
Ref.~\refcite{Dimopoulos_etal(2012)vFd-eta}.

Another important consequence of the inflationary $\mathcal{V}_{+}$
attractor solution is the scaling of the gauge kinetic function. Using
Eqs.~(\ref{x-def}) and (\ref{eV-ef-def}), we can write 
\begin{equation}
\frac{\dot{f}}{Hf}=\ef\x.\label{df}
\end{equation}
Plugging in the value of $\x_{\mathcal{V}}$ from Table~1 and using
$\left|\ef\right|\gg\left|\ev\right|$ and $\left|\ef\right|\gg1$,
we find $\ef\x_{\mathcal{V}}\simeq-2$ and therefore
\begin{equation}
f\propto a^{-2}.\label{f-scaleV}
\end{equation}
As shown in Refs.~\refcite{Dimopoulos2007} and \refcite{Dimopoulos_us(2009)_fF2}
such scaling produces a flat vector field perturbation spectrum. 

$f$ scales as a negative power of $a$ in the whole parameter region
where $\mathcal{V}_{+}$ is an attractor (the only attractor with
a non-vanishing vector field). That is, $f$ can only decrease in
time. This result is important if the current setup is to be used
in constructing models motivated by particle physics theories, where
$\a_{\mu}$ is a gauge field. In that case the inverse of the gauge
kinetic function $f^{-1}$ can be interpreted as the time-dependent
gauge coupling ``constant''\cite{Demozzi(2009)PMFs} and time-dependent
self-coupling ``constant'' in the case of non-Abelian $\a_{\mu}$.\cite{Karciauskas(2012)nAb1} Once the inflaton is stabilised, $f$
is constant, and thus can be absorbed into the definition of the gauge
field $\a_{\mu}$. Therefore, without the loss of generality one can
normalise $f_{\mathrm{end}}=1$. If $n<0$ where $f\propto a^{n}$,
the gauge kinetic function $f\gg1$ during inflation and all couplings
of the (canonically normalised) gauge field are exponentially suppressed,
making it virtually a free Abelian field. In the opposite regime $n>0$
and $f\ll1$, which makes all coupling constants exponentially large.
In that case the perturbative quantum field theory breaks down and
we can no longer trust our computations.\cite{Demozzi(2009)PMFs}
But as shown above, only $n<0$ is possible in the regime with non-vanishing
vector field, which does not present the strong coupling problem.

The spatial anisotropy does not vanish in $\mathcal{V}_{+}$ attractor
too. We can write it as
\begin{equation}
\Sigma_{\mathcal{V}}=\frac{6-\ev\left(\ef+\ev\right)}{3\ev\left(\ef+\ev\right)}\epsilon_{H\mathcal{V}},\label{SigV-gen}
\end{equation}
where $\epsilon_{H\mathcal{V}}$ is given in Eq.~(\ref{eHV-lim}).
An analogous relation was first noticed in Ref.~\refcite{Watanabe_Kanno_Soda(2009)}
in the context of their model and in the case of constant $\ev$ and
$\ef$ it was given in Refs.~\refcite{Kanno_etal(2010)attractor} and \refcite{Dimopoulos_Wagstaff(2010)BackReact}.
Looking at Eq.~(\ref{SV-bndr}), we see that close to the $\mathcal{S}$/$\mathcal{V}_{+}$
bifurcation curve, the factor in front of $\epsilon_{H\mathcal{V}}$
in Eq.~(\ref{SigV-gen}) vanishes. Further away from that curve
the second term in the numerator dominates and the spatial anisotropy
becomes
\begin{equation}
\Sigma_{\mathcal{V}}\simeq-\frac{1}{3}\epsilon_{H\mathcal{V}}.\label{SigV-lim}
\end{equation}
Hence close to the $\mathcal{S}$ region in the parameter space, the
spatial anisotropy can be arbitrarily small; otherwise it is of the
order of slow-roll parameter.

By definition, the square of expansion rate normalised variables in
Eqs.~(\ref{x-def}) and (\ref{Sig2-def}) are equal to their fractional
contribution to the total energy density of the
universe (c.f. Eq.~(\ref{constr})). Hence from Eq.~(\ref{zV-lim})
we find
\begin{equation}
\z_{\mathcal{V}}^{2}\simeq\frac{\ef\ev-6}{\ef^{2}}\ll1.
\end{equation}
That is, the vector field energy density gives a constant but subdominant
contribution to the total energy budget in the inflationary part of
the parameter space. On the other hand, the scalar field potential
energy $\y_{\mathcal{V}}^{2}\simeq1$ dominates in the same paremeter
region. More generally, we can find
\begin{equation}
\left(\frac{\z_{\mathcal{V}}}{\y_{\mathcal{V}}}\right)^{2}=\frac{\ev\left(\ef+\ev\right)-6}{\ef\left(\ef+\ev\right)+6},
\end{equation}
which is valid in the whole $\mathcal{V}_{+}$ region.

\subsection{Models with varying $\protect\ev$ and $\protect\ef$}

In the analysis above $\ev$ and $\ef$ were assumed constant.
As the slow-roll parameter $\epsilon_{H}$ is a function of these
parameters only, it also is constant. Thus the model does not provide a 
graceful exit that is, a way to end inflation. However, this is not a big
problem as one expects that at some critical value $\s_{\mathrm{c}}$
the potential of the scalar field can be modified by other degrees
of freedom in the theory which end inflation. Such a scenario could
be realised similarly to the hybrid inflation, for example. Another
consequence of constant $\epsilon_{H}$ is that the scale factor grows
as $a\propto t^{1/\epsilon_{H}}$. A power law inflation driven by
a single scalar field is excluded by observations.\cite{Ade(2015)Planck-infl}
This does not, however, guarantee that power law models discussed
in this work are all excluded. Their predicted spectral index might
be modified due to the presence of the vector field.\cite{Dimopoulos_etal(2012)vFd-eta}

Constant $\ev$ and $\ef$ parameters imply that the scalar field
potential and the gauge kinetic function are both exponential functions
of $\s$. Indeed, from the definitions in Eqs.~(\ref{eV-ef-def})
we find $V=V_{0}\exp\left[\sqrt{\frac{2}{3}}\ev\s/\mpl\right]$ and
$f=f_{0}\exp\left[\frac{1}{\sqrt{6}}\ef\s/\mpl\right]$. The time dependence
of $\ev$ or $\ef$ implies the departure from purely exponential forms
of $V\left(\s\right)$ and $f\left(\s\right)$. But even in this case
the dynamical analysis presented in this work can still be applicable,
subject to additional constraints. If $\ev$ or $\ef$ change
with time, the future asymptotic solution of Eqs.~(\ref{EoM-Sii})--(\ref{EoM-z})
will no longer be determined by a point in the plots of Figure~\ref{fig:stab-regs}.
Rather, it will be a trajectory on those plots. Starting from initial
values $\left(\ev_{*},\ef_{*}\right)$ the position of the $\left(\ev,\ef\right)$
point in the plane will change. If the change is not too fast, the
solution of Eqs.~(\ref{EoM-Sii})--(\ref{EoM-z}) will follow the
attractor corresponding to that point. The conditions which $\ev$
and $\ef$ have to satisfy were discussed in Ref.~\refcite{Dimopoulos_Wagstaff(2010)BackReact}.

As an example, consider a model presented in Ref.~\refcite{Watanabe_Kanno_Soda(2009)}.
The scalar field potential in that work is taken to be of chaotic
type $V\left(\s\right)=m^{2}\s^{2}/2$ and the gauge kinetic function
given by $f\left(\s\right)=\exp\left(c\s^{2}/2\mpl^{2}\right)$. In
this case we find
\begin{equation}
\ev=\sqrt{6}\frac{\mpl}{\s}\quad\mathrm{and}\quad\ef=c\sqrt{6}\frac{\s}{\mpl}.\label{eV-Soda}
\end{equation}
Let us take $c=1$, which is also used in Ref.~\refcite{Bartolo(2013)fF2}.
Assuming that inflation proceeds in the $\mathcal{V}_{+}$ attractor
and choosing that cosmological scales exit the horizon $N_{*}=50$
$e$-folds before the end of inflation we find $\s_{*}/\mpl\simeq14$.
Plugging this result into Eqs.~(\ref{eV-Soda}) we get  $\ev_{*}\simeq0.175$
and $\ef_{*}\simeq34.3$. The value of $\ef_{*}$ is barely larger
than the bound in Eq.~(\ref{SV-bndr}). It is thus consistent with
the assumption that inflation proceeds in the $\mathcal{V}_{+}$ attractor
but it is very close to the $\mathcal{S}$ attractor. The inflationary
trajectory in the $\left(\ev,\ef\right)$ plane of this model is represented
by the short orange curve on the right-hand side of Figure~\ref{fig:stab-regs}.
The orange dot marks the $\left(\ev_{*},\ef_{*}\right)$ values.

More variants of $V\left(\s\right)$ and $f\left(\s\right)$, which
give varying $\ev$ and $\ef$ parameters, are presented in Ref.~\refcite{Dimopoulos_Wagstaff(2010)BackReact}.
When building these types of models one should also note another subtlety.
By looking at Eq.~(\ref{SigV-lim}), we see that if the graceful exit
is realised while the dynamics of the system is determined by the
$\mathcal{V}_{+}$ attractor, the anisotropy of the universe increases
towards the end of inflation until it becomes highly anisotropic $\left|\Sigma_{\mathcal{V}}\right|\simeq1$.
Hence one needs to study how such a large anisotropy affects perturbations
and also to ensure isotropisation of the universe in the post-inflationary
period. However, this is not a concern if one introduces additional
degrees of freedom to provide the graceful exit. In that case inflation
can be terminated before $\epsilon_{H\mathcal{V}}$, and hence $\Sigma_{\mathcal{V}}$,
become large. 

\section{Conclusions}

In this work I present the dynamical analysis of a scalar-vector system
which interacts via an $f^{2}\left(\s\right)F_{\mu\nu}F^{\mu\nu}$ term.
The analysis is performed in full phase space, without restricting
initial conditions to special alignments or additional symmetries
apart from the Bianchi I spatial metric. The parameter space for each
attractor region is shown in Figure~\ref{fig:stab-regs}. It is demonstrated
that a unique attractor exists for each value of $\ev$ and $\ef$.
Bifurcation curves, where attractor points change their stability,
are clearly delineated and their full analytic expressions are given
in Eqs.~(\ref{KV-bndr})--(\ref{SK-bndr}). The stability of each
fixed point is determined by the eigenvalues of the matrix $\mathcal{M}$.
Analytic expressions of those eigenvalues are presented in Eqs.~(\ref{m1})--(\ref{m9})
in full generality, without any approximations.

The parameter space, where potentially successful models of inflation
can be build, are shown by the darker region in the right-hand side plot of
Fig.~\ref{fig:stab-regs}. The analysis is performed for constant
$\ev$ and $\ef$, that is, for the exponential scalar field potential
$V\left(\s\right)$ and kinetic function $f\left(\s\right)$. However,
stability regions represented in Fig.~\ref{fig:stab-regs} also apply
if $\ev$ and $\ef$ vary slowly, as formulated in Ref.~\cite{Dimopoulos_Wagstaff(2010)BackReact}.
In that case parameters draw a trajectory in the $\left(\ev,\ef\right)$
plane as inflation proceeds. As an example, I show the trajectory
of the model presented in Ref.~\refcite{Watanabe_Kanno_Soda(2009)}.
A more rigorous study of allowed trajectories is left for future publications.

In this letter I indicate only the parameter range where $\epsilon_{H}$
of a given inflationary attractor is within the observationally allowed
range. However, to find a realistic range, one has to compute the
values of other observables. The model of Ref.~\refcite{Watanabe_Kanno_Soda(2009)},
for example, is tightly constrained by the bounds on statistical anisotropy.\cite{Bartolo(2013)fF2,Lyth(2013)f2F2} Such analysis is beyond the
scope of this work.


\paragraph*{Acknowledgements. ~}

I would like to thank Jacques Wagstaff for numerous very long discussions
which helped to understand a lot of the issues related to this work.
I am also very grateful to Konstantinos Dimopoulos for many discussions
about the role of vector fields during inflation. Most of the results
reported in this letter were first presented in the workshop on ``Inflation
and the Origin of the CMB Anomalies'', held in Universidad del Valle,
Colombia on May 18-22, 2015. I am very grateful to organisers for
their hospitality and a perfectly organised meeting. Part of this work
was supported by JSPS during my fellowship as an International Research
Fellow of the Japan Society for the Promotion of Science. It is also
supported by the Academy of Finland project 278722.

\appendix

\section{Linearised Dynamical Equations\label{sec:Lin-EoMs}}

Using the procedure described in the main text to linearise Eqs.~(\ref{EoM-Sii})--(\ref{EoM-z}),
we obtain
\begin{eqnarray*}
\frac{\mathrm{d}\delta\Sigma_{\ii}}{\mathrm{d}N} & = & \left(\epsilon_{H}-3\right)\delta\Sigma_{\ii}+2\Sigma_{\ii}\delta\epsilon_{H}-4\left(\Sigma_{1}\delta\Sigma_{1}+\Sigma_{2}\delta\Sigma_{2}\right)-8\z\delta\z_{\ii}\\ \\
\frac{\mathrm{d}\delta\Sigma_{1}}{\mathrm{d}N} & = & \left(\epsilon_{H}-3\right)\delta\Sigma_{1}+2\Sigma_{1}\delta\epsilon_{H}+\frac{\Sigma_{\ii}\delta\Sigma_{1}+\Sigma_{1}\delta\Sigma_{\ii}}{2}-\\&&\frac{\Sigma_{+}\delta\Sigma_{1}+\Sigma_{\times}\delta\Sigma_{2}+\Sigma_{1}\delta\Sigma_{+}+\Sigma_{2}\delta\Sigma_{\times}}{\sqrt{2}}-6\delta\z_{1}\quad\quad\\ \\
\frac{\mathrm{d}\delta\Sigma_{2}}{\mathrm{d}N} & = & \left(\epsilon_{H}-3\right)\delta\Sigma_{2}+2\Sigma_{2}\delta\epsilon_{H}+\frac{\Sigma_{\ii}\delta\Sigma_{2}+\Sigma_{2}\delta\Sigma_{\ii}}{2}-\\&&\frac{\Sigma_{\times}\delta\Sigma_{1}-\Sigma_{+}\delta\Sigma_{2}+\Sigma_{1}\delta\Sigma_{\times}-\Sigma_{2}\delta\Sigma_{+}}{\sqrt{2}}-6\delta\z_{2}\\ \\
\frac{\mathrm{d}\delta\Sigma_{+}}{\mathrm{d}N} & = & \left(\epsilon_{H}-3\right)\delta\Sigma_{+}+2\Sigma_{+}\delta\epsilon_{H}+2\sqrt{2}\left(\Sigma_{1}\delta\Sigma_{1}-\Sigma_{2}\delta\Sigma_{2}\right)\\ \\
\frac{\mathrm{d}\delta\Sigma_{\times}}{\mathrm{d}N} & = & \left(\epsilon_{H}-3\right)\delta\Sigma_{\times}+2\Sigma_{\times}\delta\epsilon_{H}+2\sqrt{2}\left(\Sigma_{2}\delta\Sigma_{1}+\Sigma_{1}\delta\Sigma_{2}\right)\\ \\
\frac{\mathrm{d}\delta\x}{\mathrm{d}N} & = & \left(\epsilon_{H}-3\right)\delta\x+2\x\delta\epsilon_{H}+2\ev\left(\x\delta\x+\Sigma\delta\Sigma\right)+2\left(\ef+\ev\right)\z\delta\z_{\ii}\\ \\
\frac{\mathrm{d}\delta\z_{\ii}}{\mathrm{d}N} & = & \left(\epsilon_{H}-3\right)\delta\z_{\ii}+2\z\delta\epsilon_{H}-\left(\ef\x-1-\Sigma_{\ii}\right)\delta\z_{\ii}-\z\left(\ef\delta\x-\delta\Sigma_{\ii}\right)\\ \\
\frac{\mathrm{d}\delta\z_{1}}{\mathrm{d}N} & = & \left(\epsilon_{H}-3\right)\delta\z_{1}-\left(\ef\x-1-\frac{5}{2}\Sigma_{\ii}\right)\delta\z_{1}-\frac{\Sigma_{+}\delta\z_{1}+\Sigma_{\times}\delta\z_{2}}{\sqrt{2}}\\&&+2\left(\z\delta\Sigma_{1}+\Sigma_{1}\delta\z_{\ii}\right)\\ \\
\frac{\mathrm{d}\delta\z_{2}}{\mathrm{d}N} & = & \left(\epsilon_{H}-3\right)\delta\z_{2}-\left(\ef\x-1-\frac{5}{2}\Sigma_{\ii}\right)\delta\z_{2}-\frac{-\Sigma_{+}\delta\z_{2}+\Sigma_{\times}\delta\z_{1}}{\sqrt{2}}\\&&+2\left(\z\delta\Sigma_{2}+\Sigma_{2}\delta\z_{\ii}\right)\qquad\quad
\end{eqnarray*}
where I used the notation 
\begin{eqnarray*}
\delta\epsilon_{H} & \equiv & 2\left[3\left(\x\delta\x+\Sigma\delta\Sigma\right)+\z\delta\z_{\ii}\right],\\
\Sigma\delta\Sigma & \equiv & \frac{1}{6}\left(\frac{3}{2}\Sigma_{\ii}\delta\Sigma_{\ii}+2\Sigma_{1}\delta\Sigma_{1}+2\Sigma_{2}\delta\Sigma_{2}+\Sigma_{+}\delta\Sigma_{+}+\Sigma_{\times}\delta\Sigma_{\times}\right)
\end{eqnarray*}

Formally these equations can be written as in Eq.~(\ref{diff-eqns-formal}).
The eigenvalues of the matrix $\mathcal{M}$ for each fixed
point in Table~1 can be compactly expressed as

\begin{eqnarray}
m_{1,2,3} & = & \epsilon_{H}-3\label{m1}\\
m_{4} & = & \epsilon_{H}-3-\frac{1}{2}\left[\Delta_{a}-\sqrt{\Delta_{a}^{2}-8\left(6+\ef\left(\ef+\ev\right)\right)\z^{2}}\right]+2\left(\epsilon_{H}+\ev\x\right)\\
m_{5} & = & \epsilon_{H}-3-\frac{1}{2}\left[\Delta_{a}+\sqrt{\Delta_{a}^{2}-8\left(6+\ef\left(\ef+\ev\right)\right)\z^{2}}\right]\\
m_{6,7} & = & \epsilon_{H}-3-\frac{1}{2}\left[\Delta_{b}-\Sigma_{\ii}-\sqrt{\Delta_{b}^{2}-48\z}\right]\\
m_{8,9} & = & \epsilon_{H}-3-\frac{1}{2}\left[\Delta_{b}-\Sigma_{\ii}+\sqrt{\Delta_{b}^{2}-48\z}\right]\label{m9}
\end{eqnarray}
where $\Delta_{a}\equiv\ef\x-1-\Sigma_{\ii}$ and $\Delta_{b}\equiv\ef\x-1-2\Sigma_{\ii}$.
To find explicit expressions for each fixed point, one has to
plug in the values of $\Sigma_{\ii}$, $\x$, $\z$ and $\epsilon_{H}$
corresponding to that point and take the positive branch $\sqrt{\Delta^{2}}=+\Delta$
where appropriate.


\bibliographystyle{ws-mpla}
\bibliography{}

\end{document}